\documentclass[10pt]{article}
\pdfoutput=1
\usepackage{amssymb,amsmath,mathrsfs,enumerate}
\usepackage{graphicx,rotate,multicol,slashed}
\usepackage{wrapfig}
\usepackage[dvipsnames]{xcolor}
\definecolor{mygreen}{RGB}{06,84,39}
\usepackage[colorlinks=true,
            linkcolor=mygreen,
            urlcolor=mygreen,
            citecolor=mygreen]{hyperref}
\usepackage{soul}
\usepackage{cite}
\usepackage{verbatim}
\usepackage[symbol]{footmisc}
\renewcommand*{\thefootnote}{\fnsymbol{footnote}}

\long\def\rpl#1!!#2!!{\textcolor{red}{#1} \textcolor{blue}{#2}}
\newcommand{\hc}[0]{\mathrm{h.c.}}

\def \order(#1){{\cal O} \left(#1 \right)}

\evensidemargin=\oddsidemargin
\def\Eqn#1{Eq.\ (\ref{#1})}

\allowdisplaybreaks
\begin{document}
%
%\begin{flushright}
%    LU TP 18-35\\
%    TIFR/TH/19-xx
%\end{flushright}

\begin{center}
	{\Large \bf Complementary bound on $W^\prime$ mass from Higgs to diphoton decay } \\
	\vspace*{1cm} {\sf Triparno Bandyopadhyay$^{a,}
        $\footnote{\href{mailto:triparno@theory.tifr.res.in}{\texttt{triparno@theory.tifr.res.in}}},~
        Dipankar
        Das$^{b,}$\footnote{\href{mailto:dipankar.das@thep.lu.se}{\texttt{dipankar.das@thep.lu.se}}},~Roman
        Pasechnik$^{b,}$\footnote{\href{mailto:roman.pasechnik@thep.lu.se}{\texttt{roman.pasechnik@thep.lu.se}}},~Johan 
    Rathsman$^{b,}$\footnote{\href{mailto:johan.rathsman@thep.lu.se}{\texttt{johan.rathsman@thep.lu.se}}}} \\
	\vspace{10pt} {\small \em 
    %%%%%%%%%%%%%%%%%%%%%%%%%%%%%%%%%%%%%%%%%%%%%%%%%%%%%%%%%%%%%
          $^a$Department of Theoretical Physics,Tata
          Institute of Fundamental Research, Mumbai 400005, 
          India \\
	% % % % % % % % % % % % % % % % % % % % % % % % % % % % % % % % % % % % % %
	      $^b$Department of Astronomy and Theoretical Physics, Lund
          University, S\"{o}lvegatan 14A, 223 62 Lund, Sweden
    % % % % % % % % % % % % % % % % % % % % % % % % % % % % % % % % % % % % % %
 }
	
	\normalsize
\end{center}

%%%%%%%%%%%%%%%%%%%%%%%%%%%%%%%%%%%%%%%%%%%%%%%%%%%%%%%%%
%%%%  Changing footnote label back to arabic numbers %%%%
%%%%%%%%%%%%%%%%%%%%%%%%%%%%%%%%%%%%%%%%%%%%%%%%%%%%%%%%
\renewcommand*{\thefootnote}{\arabic{footnote}}
%%%%%%%%%%%%%%%%%%%%%%%%%%%%%%%%%%%%%%%%%%%%%%%%%%%%%%%
%%%%%  Resetting footnote counter  %%%%%%%%%%%%%%%%%%%
%%%%%%%%%%%%%%%%%%%%%%%%%%%%%%%%%%%%%%%%%%%%%%%%%%%%%%
\setcounter{footnote}{0}

\begin{abstract}
    \noindent
    Using the left-right symmetric model as an illustrative example,
    we suggest a simple and straightforward way 
%    of probing the scale of
%   left-right symmetric gauge theories and physics of the corresponding
%    $W'$ bosons by means 
of constraining the $W'$ mass directly from
    the decay of the Higgs boson to two photons. The proposed method is
    generic and applicable to a diverse range of models with a
    $W'$-boson that couples to the SM-like Higgs boson.  Our analysis
    exemplifies how the precision measurement of the Higgs to diphoton
    signal strength can have a pivotal role in probing the scale of new
    physics.
\end{abstract}

%\section{Introduction}
%
Models that extend the standard electroweak~(EW) gauge symmetry,
$\mathcal{G}_{\rm EW}\sim SU(2)_L\times U(1)_Y$, to a larger group,
$\mathcal{G}'$, often end up introducing new, electrically charged gauge
bosons.  The left-right symmetric model\cite{lrs1, lrs2, lrs3, lrs4}
where $\mathcal{G}'$ is identified with the gauge group $SU(2)_L\times
SU(2)_R \times U(1)_X$, constitutes a well-motivated example of such a
framework. In this model, $W_L^\pm$ and $W_R^\pm$, the charged gauge
bosons corresponding to $SU(2)_L$ and $SU(2)_R$ respectively, would mix
to produce the physical eigenstates $W^\pm$ and $W^{\prime\pm}$ as
follows
\begin{subequations}
    \label{e:wlwr}
    \begin{eqnarray}
    W^\pm &=& \cos\xi \, W_L^\pm + \sin\xi \, W_R^\pm \,, \\ 
    W^{\prime\pm} &=& -\sin\xi \, W_L^\pm + \cos\xi \, W_R^\pm \,,
    \end{eqnarray}
\end{subequations}
where $W$ is assumed to be the lighter mass eigenstate later to be
identified with the $W$-boson of the Standard Model~(SM).  Due to such a
mixing, the tree-level value of the EW $\rho$-parameter as well as the
$W$-boson couplings are shifted from their corresponding SM
expectations. The existing EW precision data restricts the mixing angle
to be very small ($\xi < 10^{-2}$)\cite{PDG}.

Considerable efforts have been made to look for such heavy
$W^\prime$-bosons via direct and indirect searches. Non-observation of
any convincing signature has led to lower bounds on the mass of
the $W^\prime$-boson~($M_{W'}$). Indirect bounds on $M_{W'}$ have been
placed using many different considerations such as 
Michel parameters ($M_{W'}>250$~GeV from muon decay and 
$M_{W'}>145$~GeV from tauon decay)
\cite{Prieels:2014paa,Ackerstaff:1998yk}, parity violation in polarized
muon decays ($M_{W'}>600$~GeV)\cite{Bueno:2011fq}, neutral meson
oscillations ($M_{W'}>2.5$~TeV) \cite{Gaillard:1974hs, Beall:1981ze,
Bertolini:2014sua}, $CP$-violating observables in Kaon decay
($M_{W'}>4.2$~TeV) \cite{Zhang:2007da}, and the neutron electric dipole
moment ($M_{W'}>8$~TeV) \cite{Zhang:2007da}. All these bounds rely
heavily on the fermionic couplings of the $W^\prime$-boson.
Additionally, the constraints arising from the observables involving the
quark sector depend on the right-handed CKM matrix which is usually
presupposed to be equal to its left-handed counterpart.  Quite
unsurprisingly, all these bounds can be diluted substantially once 
the assumptions about the fermionic
couplings are relaxed\cite{Langacker:1989xa,Datta:1982ce,Datta:1982ek,Basak:1983ft}.
Direct searches for $W^\prime$ have also been performed at the
LHC in a plethora of final states\cite{Sirunyan:2016iap,
    Aaboud:2017yvp,Khachatryan:2015pua,Aaboud:2017efa,
    Khachatryan:2016jww, Aaboud:2017ahz,Sirunyan:2017wto,
    Khachatryan:2014dka,Aad:2015xaa} with bounds in the few TeV range. These searches, again,
rely on assumptions about the branching ratios~(BRs) of $W^{\prime}$
into different channels, which, in turn, depend on the fermionic
couplings of the $W'$-boson.

In this paper we, on the other hand, make an effort to place bound on
$M_{W'}$ without appealing at all to the fermionic couplings of the
$W^\prime$-boson. Evidently, such a bound would go well beyond the ambit
of left-right symmetry and will be applicable to a much wider
variety of $SU(2)\times SU(2)\times U(1)$ models \cite{Hsieh:2010zr}.
Our strategy is based on the realization that very often the
$W^{\prime}$-boson receives part of its mass from the vacuum expectation values~(VEVs) at the EW
scale. Consequently, the SM-like scalar~($h$) observed at the LHC, which
must somehow emerge from the scalar sector of the extended gauge theory,
should possess trilinear coupling of the form $W^\prime W^\prime h$ with
strength proportional to the fraction of $M_{W'}$ that stems from the EW
scale VEVs. It is this `fraction' which can be sensed via the precision
measurement of the Higgs to diphoton signal strength. In anticipation
that the Higgs signal strengths will continue to agree with the
corresponding SM expectations with increasing accuracy, we should be
able to estimate how heavy the $W^\prime$-boson needs to be compared to
the EW scale.
Before moving on to the main part, let us brief the key assumptions
that enter our analysis:
\begin{enumerate}[(i)]
    \item The $W$-$W^\prime$ mixing is very small ($\xi \to 0$),
    which, in the context of left-right symmetry, 
    is consistent with the fact that
    the charged currents mediated by the $W$-boson at low energies
    are mostly left-handed.
    \item An SM-like Higgs scalar, $h$, emerges as a linear
    combination of the components of the scalar fields present in
    the theory. In view of the current Higgs data\cite{CMS:2018lkl}, 
    this is a     reasonable assumption.
    \item The physical charged scalars are heavy enough to have
    essentially decoupled from the EW scale observables. Therefore,
    the $W^\prime$-boson will give the dominant new physics~(NP)
    contribution to the Higgs to diphoton decay amplitude.
\end{enumerate}

%%%%%%%%%%%%%%%%%%%%%%%%%%
%\section{Setting up the formalism}
%\label{sec:formalism}
%%%%%%%%%%%%%%%%%%%%%%%%%%

To illustrate the idea further, we consider the example of a 
left-right symmetry which is broken spontaneously 
by the following scalar multiplets:
\begin{eqnarray}
\label{e:transformation}
    \phi \equiv (2,2,x_\phi) \,, \quad
    \chi_L \equiv (2,1,x_L) \,, \quad
    \chi_R \equiv (1,2,x_R) \,,
\end{eqnarray}
where the quantities inside the brackets characterize the transformation
properties under the gauge group $SU(2)_L\times SU(2)_R\times U(1)_X$.\footnote{
	In more conventional left-right symmetric models $\chi_L$ and $\chi_R$ are triplets
	of $SU(2)_L$ and $SU(2)_R$ respectively. In these cases, however, the VEV of $\chi_L$
	has to be smaller than $\order(1~{\rm GeV})$\cite{Blank:1997qa,Akeroyd:2007zv,Kanemura:2012rs,Das:2016bir} so that the tree-level value of the
	EW $\rho$-parameter is not substantially altered from unity.}
Note that, the main analysis of our paper will not depend on the $U(1)_X$ charge assignments.
After the spontaneous symmetry breaking (SSB), the scalar multiplets 
are expanded as follows:
\begin{eqnarray}
\label{e:SSB}
    \phi = \frac{1}{\sqrt{2}} \begin{pmatrix}
        v_1+h_1+i z_1 & \sqrt{2} w_2^+ \\
        \sqrt{2} w_1^- & v_2+h_2+i z_2
    \end{pmatrix}\,, \quad % \\
    \chi_L = \frac{1}{\sqrt{2}} \begin{pmatrix}
      \sqrt{2}  w_L^+ \\ v_L+h_L+i z_L
    \end{pmatrix} \,, \quad
    \chi_R = \frac{1}{\sqrt{2}} \begin{pmatrix}
      \sqrt{2}  w_R^+ \\ v_R+h_R+i z_R
    \end{pmatrix} \,,
\end{eqnarray}
where $v_i$ $(i=1,2)$, $v_L$ and $v_R$ denote the VEVs of $\phi$,
$\chi_L$ and $\chi_R$, respectively. The kinetic terms for the
 scalar sector reads
\begin{eqnarray}
\label{e:skin}
  {\mathscr L}_{\rm kin} = {\rm Tr}[(D_{\mu}\phi)^\dagger(D^{\mu}\phi)] 
  + (D_{\mu}\chi_{L})^\dagger(D_{\mu}\chi_{L}) 
    + (D_{\mu}\chi_{R})^\dagger(D_{\mu}\chi_{R}) \,,
\end{eqnarray}
where the covariant derivatives are given by
\begin{subequations}
    \label{e:cov}
    \begin{eqnarray}
    D_{\mu}\phi &=& \partial_\mu \phi + i  (g_L W_{\mu L}\phi - g_R \phi W_{\mu R}) +i g_x x_\phi \phi \,, \\
    D_{\mu}\chi_{L(R)} &=& \partial_\mu \chi_{L(R)} + i g_{L(R)} W_{\mu L(R)} \chi_{L(R)} + ig_x x_{L(R)} X_\mu \,.
    \end{eqnarray}
\end{subequations}
In the above equations, the quantities $g_{L(R)}$ and $g_x$ represent
the gauge coupling strengths corresponding to $SU(2)_{\rm L(R)}$ and
$U(1)_X$ respectively whereas $X_\mu$ stands for the gauge field
corresponding to $U(1)_X$.
The $SU(2)_{\rm L(R)}$ gauge fields can be conveniently expressed in
the matrix form as
\begin{eqnarray}
    W_{\mu L(R)} \equiv \frac{\sigma_a}{2} W_{\mu L(R)}^a = \frac12
    \begin{pmatrix}
       W_{\mu L(R)}^3 & \sqrt{2}W_{\mu L(R)}^+ \\
        \sqrt{2}W_{\mu L(R)}^- & -W_{\mu L(R)}^3
    \end{pmatrix}\,.
\end{eqnarray}
 In what follows, we are interested only in the charged components 
 $W_{\mu L(R)}^\pm$. The corresponding mass squared matrix in the
  $W_{L}$-$W_{R}$ basis is found to be
\begin{eqnarray} 
\label{e:MLR}
    {\cal M}_{\rm LR}^2 
    = \frac{1}{4}
    \begin{pmatrix}
        g_L^2 (v_1^2 + v_2^2 + v_L^2) & -2 g_L g_R v_1v_2 \\
        -2 g_L g_R v_1v_2 & g_R^2 (v_1^2 + v_2^2 + v_R^2)
    \end{pmatrix}\,.
\end{eqnarray}
This mass squared matrix can be diagonalized by the orthogonal
rotation given in \Eqn{e:wlwr}. This rotation will then entail the
following relations:
\begin{subequations}
\label{e:mass}
\begin{eqnarray}
    M_W^2 \cos^2\xi +M_{W'}^2 \sin^2\xi &=& \frac{g_L^2}{4}
    (v_1^2+v_2^2+v_L^2) \,, \label{e:mass1} \\
    M_W^2 \sin^2\xi +M_{W'}^2 \cos^2\xi &=& \frac{g_R^2}{4}
    (v_1^2+v_2^2+v_R^2) \,, \label{e:mass2} \\
    \left(M_{W'}^2 -M_W^2\right) \sin\xi \cos\xi &=&
    \frac{g_L g_R}{2} v_1 v_2 \,.
\end{eqnarray}
\end{subequations}
In the limit $\xi \to 0$ we can rewrite \Eqn{e:mass1} as
\begin{eqnarray}
    \label{e:wmass}
    M_W^2 \approx \frac{g_L^2}{4}(v_1^2+v_2^2+v_L^2) \equiv
    \frac{g_L^2 v^2}{4} \,,
\end{eqnarray}
where, we have identified the EW VEV as
\begin{eqnarray}
    \label{e:ewvev}
    v=\sqrt{v_1^2+v_2^2+v_L^2} = 246~{\rm GeV} \,.
\end{eqnarray}
At this point, let us define the SM-like Higgs scalar as 
follows:\footnote{
    We are implicitly assuming that the parameters in the scalar
    potential are adjusted properly so that $h$ becomes a physical
    eigenstate.}
\begin{eqnarray}
\label{e:h}
h = \frac{1}{v}(v_1 h_1 + v_2 h_2 + v_L h_L) \,,
\end{eqnarray}
where $h_{1,2,L}$ are the component fields defined in \Eqn{e:SSB}. To
convince ourselves that the couplings of $h$ are indeed SM-like,
it is instructive to look at the trilinear gauge-Higgs couplings
which stem from the scalar kinetic terms of \Eqn{e:skin}. We notice
that
\begin{eqnarray}
    {\mathscr L}_{\rm kin} \ni \frac{g_L^2}{2} W_{\mu L}^+ W_L^{\mu -}
    (v_1h_1 +v_2h_2 +v_Lh_L) = \frac{g_L^2 v}{2} W_{\mu L}^+ W_L^{\mu -}
    h \,.
\end{eqnarray}
Since in the limit $\xi \to 0$ the $W$-boson almost entirely overlaps
with $W_L$, following \Eqn{e:wmass}, we can rewrite the above equation
as
\begin{eqnarray}
{\mathscr L}_{\rm kin} \ni g_L M_W \, W_{\mu}^+ W^{\mu -} h \,.
\end{eqnarray}
Clearly, the tree-level $WWh$ coupling is exactly SM-like.\footnote{
    Similarly, to ensure that the tree-level $ZZh$ coupling is also
    SM-like, we would require the $Z$-$Z^\prime$ mixing in the
    neutral gauge boson sector to be small, which is sensible 
    too\cite{Czakon:1999ha,Erler:2009jh, Bandyopadhyay:2018cwu}.
    }
In the Appendix we show that the Yukawa couplings of $h$ with the SM 
fermions are also SM-like at the tree-level.

Now that we have established that $h$ possesses SM-like couplings, the
production and the tree-level decays of $h$ will remain SM-like too.
However, the loop induced decay modes such as $h\to \gamma\gamma$ will
pick up additional contributions arising from the $W^\prime$-loop. To
analyze the impact of the $W^\prime$-boson, let us first write down the
effective $h\gamma\gamma$ coupling as follows:
\begin{equation}
\label{e:hgg}
\mathscr{L}_{h\gamma\gamma} =
g_{h\gamma\gamma}F_{\mu\nu}F^{\mu\nu}h \,,
\end{equation}
where $F^{\mu\nu}=\partial^{\mu}A^{\nu} -\partial^{\nu} A^{\mu}$ is the
usual electromagnetic field tensor. Then the $h\gamma\gamma$ coupling
modifier can be defined as
\begin{equation}
\label{e:kg}
\kappa_\gamma =
\frac{g_{h\gamma\gamma}}{\left( g_{h\gamma\gamma} \right)^{\mathrm{SM}}}  \,,
\end{equation}
which, under the assumption that the $W^\prime$-boson gives the dominant
NP contribution, can be expressed as
\begin{equation}
    \label{e:kg1}
    \kappa_{\gamma} = \frac{ \mathcal{A}_1(\tau_W) + \sum_f Q_f^2
        N_c^f\mathcal{A}_{1/2} (\tau_f) + \lambda_{W'}
    \mathcal{A}_1(\tau_{W'})}{\mathcal{A}_1(\tau_W) + \sum_f Q_f^2
        N_c^f  \mathcal{A}_{1/2} (\tau_f)} \,,
\end{equation}
where $Q_f$ and $N_c^f$ stand for the electric charge and the color factor
respectively for the fermion, $f$, and, defining $\tau_a = (2m_a/m_h)^2$,
the loop functions are given by\cite{Gunion:1989we}:
\begin{subequations}
\begin{eqnarray}
    \label{e:loop}
    \mathcal{A}_1(\tau_a) &=& 2 + 3 \tau_a + 3\tau_a(2-\tau_a)f(\tau_a) \,, \\
    \mathcal{A}_{1/2}(\tau_a) &=& -2\tau_a [1+(1-\tau_a)f(\tau_a)]  \,,
\end{eqnarray}
\end{subequations}
where, for $\tau > 1$,
\begin{equation}
    \label{e:f}
    f(\tau) = \left[\sin^{-1}\left(\sqrt{\frac{1}{\tau}} \right) \right]^2 \,.
\end{equation}
The dimensionless quantity $\lambda_{W'}$ appearing in \Eqn{e:kg1} encapsulates the
contribution of the $W^\prime$-boson to the $h\to\gamma\gamma$ amplitude. In the
limit $\xi \to 0$ the expression for $\lambda_{W'}$ can be obtained as
\begin{eqnarray}
\label{e:lWp}
	\lambda_{W'} = \frac{g_{W'W'h}}{M_{W'}^2} \frac{M_W}{g_L} \approx 
	\frac{(v^2-v_L^2)}{(v^2-v_L^2+v_R^2)} \,,
\end{eqnarray}
where $g_{W'W'h}$ represents the strength of the $W^{\prime \mu} W^\prime_{\mu} h$ coupling,
which, in the limit $\xi \approx 0$, is given by
\begin{eqnarray}
	g_{W'W'h} = \frac{g_R^2}{2v} \left(v_1^2+v_2^2\right) 
	= \frac{g_R^2}{2v} \left(v^2-v_L^2\right) \,,
\end{eqnarray}
and the expression for $M_{W'}$ can be read from \Eqn{e:mass2}. The appearance
of the factor $M_W/g_L$ in \Eqn{e:lWp} is a reflection of the fact that the
quantity $g_L/M_W$ is implicitly assumed to be factored out while writing the
$h\to \gamma\gamma$ amplitude in the SM\cite{Gunion:1989we}. More interestingly
in the limit $v_L\ll v\ll v_R$, $\lambda_{W'}$ in \Eqn{e:lWp}, which
parametrizes the NP effect in $h\to\gamma\gamma$, can be approximated as
\begin{eqnarray}
\lambda_{W'} \approx \frac{v^2}{v_R^2} \,.
\end{eqnarray}
Thus, precision measurement of the $h\to \gamma\gamma$ signal strength will
be sensitive to $v_R$, {\it i.e.}, the scale of NP, irrespective of the
value of the $SU(2)_R$ gauge coupling ($g_R$), which is a clear upshot of
our analysis.

%%%%%%%%%%%%%%%%%%%%%%%%%%%%%%%%%%%%%%%%%%%%%%%%%%%%%%%%%%%%%%
%%%%%%%%%%%%%%%%%%%%%%%%%%%%%%%%%%%%%%%%%%%%%%%%%%%%%%%%%%%%%%

\begin{figure}[htpb]
    \centering
    \includegraphics[width=0.47\linewidth]{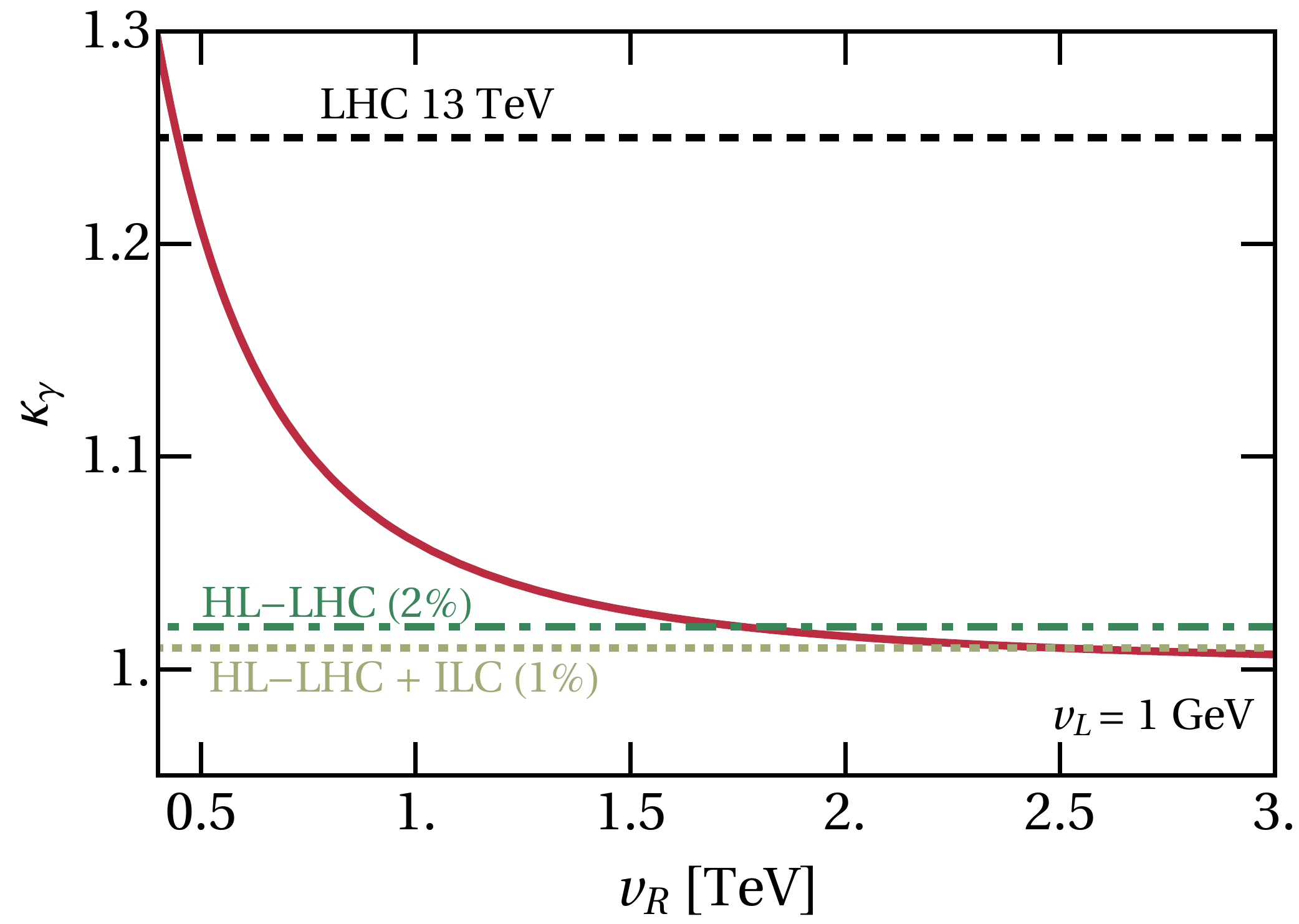}\quad 
    \includegraphics[width=0.47\linewidth]{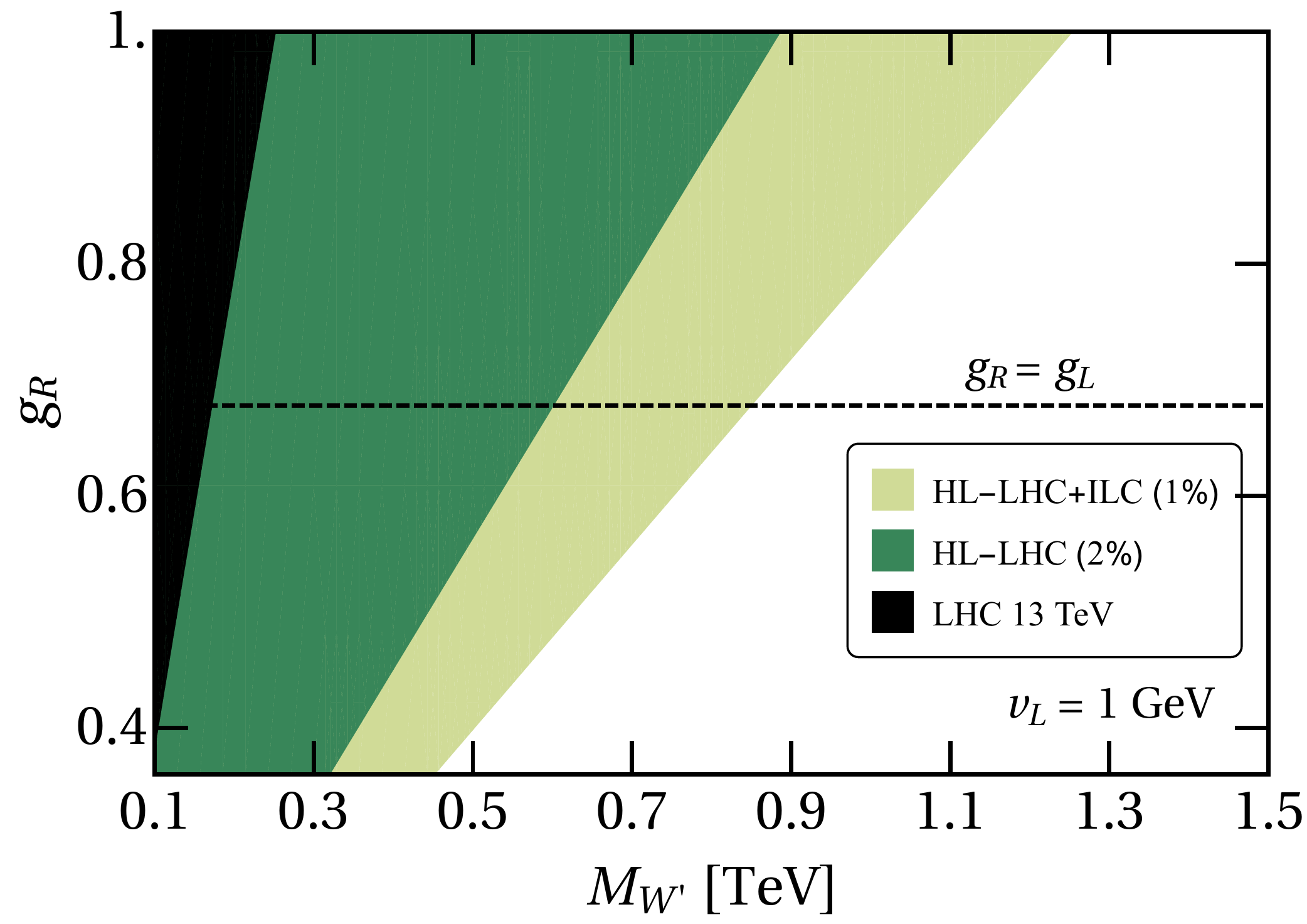}
    \caption{
            [Left Panel]{\em The solid red curve shows the variation of
                $\kappa_\gamma$, following \Eqn{e:kg1}, with $v_R$, for
                $v_L=1$ GeV. The black dashed horizontal line denotes
                the current $2\sigma$ upper limit on $\kappa_\gamma$ at
                the LHC13 ($\sqrt{s}=13$ TeV, 36 fb$^{-1}$ of data)
                \cite{CMS:2018lkl,Cepeda:2019klc}. The dark-green
                (dash-dotted) and light-green (dotted) horizontal lines
                denote the projected accuracy of on $\kappa_\gamma$ from
                the HL-LHC (2\%) data and HL-LHC+ILC (1\%) combined data
                respectively\cite{Fujii:2015jha}. Note that the
                variation in $\kappa_\gamma$ with $v_R$ and subsequently
                the limits on $v_R$ are independent of $g_R$.} [Right
                Panel] {\em The shaded area in black denotes the region
                    in the $g_R$-$M_{W'}$ plane, excluded at 95\% CL
                    from determination of $\kappa_\gamma$ at the LHC13.
                    The dark- and light-green shaded regions denote the
         excluded regions for the projected accuracy of $\kappa_\gamma$
          determination from the HL-LHC (2\%) and HL-LHC+ILC (1\%)
                    combined data respectively. While extracting bounds
                    using the projected accuracies at the HL-LHC and
                    HL-LHC+ILC, in both panels, we have assumed the
                    central value of $\kappa_\gamma$ to be unity, {\it
                i.e.}, consistent with the SM.}
        }
\label{f:kg}
\end{figure}

%%%%%%%%%%%%%%%%%%%%%%%%%%%%%%%%%%%%%%%%%%%%%%%
%%%%%%%%%%%%%%  Numerical Results  %%%%%%%%%%%%
%%%%%%%%%%%%%%%%%%%%%%%%%%%%%%%%%%%%%%%%%%%%%%%

In Fig.~\ref{f:kg} we display the bounds arising from the current as
well as future measurements of $\kappa_\gamma$. From the left panel we
can see that, irrespective of the value of $g_R$, we can rule out $v_R$ up to $450$~GeV (implying $M_{W'}\gtrsim 170$~GeV for $g_L=g_R$) at
95\% C.L. using the current LHC data\cite{CMS:2018lkl}.  Although this limit
is weak compared to the existing bounds on $M_{W'}$, it is evident from
the left panel of Fig.~\ref{f:kg} that, due to the almost horizontal
tail of the red curve, once $\kappa_\gamma$ is found to be consistent
with the SM with accuracy of a few percent at future colliders, a slight
improvement in the precision can substantially strengthen the bound on
$v_R$. To put it into perspective, 
as shown in Fig.~\ref{f:kg}, if $\kappa_\gamma$
is observed to be in agreement with the SM with a projected accuracy of
2\% at the HL-LHC\cite{Fujii:2015jha,Cepeda:2019klc}, then we can reach $v_R\gtrsim 
1.7$~TeV, which can complement the bounds from other considerations.
Furthermore, if we can attain the accuracy of 1\% in the combined
measurement of $\kappa_\gamma$ at the HL-LHC and ILC\cite{Fujii:2015jha}, then the
bound on $v_R$ can climb up to $v_R\gtrsim 2.5$~TeV.
In passing, we note that, although Fig.~\ref{f:kg} has been obtained by
setting $v_L=1$~GeV, we have checked that the plots do not crucially
depend on the exact value of $v_L$ as long as $v_L\lesssim
\order(10~{\rm GeV})$. Additionally, we have also checked that for
$v_L\lesssim \order(1~{\rm GeV})$, the constraints in Fig.~\ref{f:kg}
also apply to the more traditional versions of left-right symmetric
models where $\chi_L$ and $\chi_R$ in \Eqn{e:transformation} are
triplets of $SU(2)_L$ and $SU(2)_R$ respectively.

%%%%%%%%%%%%%%%%%%%%%%%%%%%%%%%%%%%%%%%%%%%%
%%%%%%%%%   Conclusions  %%%%%%%%%%%%%%%%%
%%%%%%%%%%%%%%%%%%%%%%%%%%%%%%%%%%%%%%%%%%%
To summarize, we have pointed out the possibility to put bound on the
mass of a $W^\prime$-boson arising from an extended gauge structure
and the corresponding symmetry breaking scale,
using an alternative set of assumptions that does not rely upon the
fermionic couplings of the $W^\prime$-boson. In view of the fact that 
the Higgs data is gradually drifting towards the SM expections with 
increasing accuracy, identifying an SM-like
Higgs boson plays an important role in our analysis. The fraction of
$M_{W'}$, that can be attributed to the EW scale, is then constrained
using the $h\to \gamma\gamma$ signal strength measurements. In our
example of a left-right symmetric scenario we find that the current data
imposes $M_{W'}\gtrsim 170$~GeV at 95\% C.L. which is at par with the
bound from the Michel
Parameters\cite{Prieels:2014paa,Ackerstaff:1998yk}, but without any
assumption about the $W'$ coupling to the right-handed leptons. One 
should also keep
in mind that the bounds from direct searches can get considerably
diluted for fermiophobic $W'$ bosons \cite{Donini:1997yu,
Chivukula:2006cg, He:1999vp, Cheng:2003ju}. Additionally, in the limit
of vanishing $W$-$W'$ mixing, the production of $W'$ via $WZ$ fusion is
also suppressed. Thus, considering the fact that the formalism described
in this paper does not depend on these factors, our bound using $h\to
\gamma\gamma$ signal strength measurements complements the existing
limits on $M_{W'}$.  Moreover, it is also encouraging to note that the
bound can rise up to $v_R> 2.5$~TeV (corresponding to $M_{W'}> 850$~GeV) if the measurement of the diphoton
signal strength is found to be consistent with the SM with a projected
accuracy of 1\% at the HL-LHC and ILC. Evidently, our current analysis
underscores the importance of the precision measurement of the
Higgs to diphoton signal strength in current as well as future collider
experiments, which can give us potential hints for the scale of NP.

%\textcolor{Brown}{To summarize, we have pointed out the possibility of bounding the mass
%of a $W^\prime$ and the corresponding symmetry breaking scale from $h\to
%\gamma\gamma$ measurements. The analysis holds for all $W'$ the mass of
%which has a contribution from the EW vev. We do not use any information
%about the fermionic couplings of the $W'$, also we take $W$-$W'$ mixing
%to be zero, rendering this bound extremely robust against model
%building. The major assumption that goes into our analysis is that the
%125 GeV scalar resonance discovered by the LHC collaborations is indeed
%the Higgs predicted by the standard model, and identifying this SM-like
%Higgs boson plays an important role in our analysis. Our computations
%shows that the current data restricts the symmetry breaking scale of new
%physics to be $\sim$ 500 GeV. It is encouraging to note that this bound
%can potentially rise to $2.5$~TeV if the diphoton signal strength is
%found to be consistent with the SM with a projected accuracy of 1\%
%using combined data of the HL-LHC and the ILC. This bound when
%translated to the mass of the $W'$, constrains it to be above 850 GeV
%($\sim$ 200 GeV from current data). Evidently, our current analysis
%underscores the importance of the precision measurement of the Higgs to
%diphoton signal strength in current as well as future collider
%experiments, which can give us potential hints for the scale of NP.}

%%%%%%%%%%%%%%%%%%%%%%%%%%%%%%%%%%%%%%%%%%%%%%%
%%%%%%%  Acknowledgments  %%%%%%%%%%%%%%%%%%%%
%%%%%%%%%%%%%%%%%%%%%%%%%%%%%%%%%%%%%%%%%%%%%%%

{\bf Acknowledgments: }{\small  D.D., R.P. and J.R. are partially supported
    by the Swedish Research Council, contract numbers 621-2013-4287 and
    2016-05996, as well as by the European Research Council (ERC) under
    the European Union's Horizon 2020 research and innovation programme
    (grant agreement No 668679).  R.P. is supported in part by CONICYT
    grant MEC80170112 as well as by the Ministry of Education, Youth and
Sports of the Czech Republic, project LT17018.}

%%%%%%%%%%%%%%%%%%%%%%%%%%%%%%%%%%%%%%%%%%%%%%%%%%%%%%%%%
%%%%%%%%%%%%%%%  Appendix  %%%%%%%%%%%%%%%%%%%%%%%%%
%%%%%%%%%%%%%%%%%%%%%%%%%%%%%%%%%%%%%%%%%%%%%%%%%%%%%%

\section*{Appendix}%
\label{sec:appendix}

%The transformation of the quark multiplets of the LR model under the
%gauge symmetry $SU(3)_C\times SU(2)_L\times SU(2)_R\times U(1)_X$ is
%given by (three generations of):
%\begin{equation}
%    q_L\equiv (3,2,1,1/3)\equiv \begin{pmatrix} u'_L \\ d'_L
%        \end{pmatrix};\; q_R\equiv (3,1,2,1/3)\equiv \begin{pmatrix}
%        u'_R\\d'_R \end{pmatrix},\label{eq:app1}
%\end{equation}
The Yukawa Lagrangian for the quark sector is given by
\begin{equation}
    \label{e:yuk}
    \mathscr{L}^{q}_{Y} = - \, \overline{Q}_L \left( Y_q\phi + 
    \widetilde{Y}_q\widetilde{\phi}\right) \, Q_R  \,,
\end{equation}
where $Q_{L(R)}=(u_{L(R)}, d_{L(R)})^T$ denotes the $SU(2)_{L(R)}$
quark doublet and we have suppressed the flavor indices. Therefore, 
$Y_q$ and $\widetilde{Y}_q$ are $3\times 3$ Yukawa matrices.
From the above Lagrangian, the
mass matrices for the up  and the
down-type quarks can be written as
\begin{equation}
    \label{e:qmass}
    M_u = \frac{1}{\sqrt{2}} \left( v_1 Y_q + v_2 \widetilde{Y}_q\right) \,, \qquad
    M_d = \frac{1}{\sqrt{2}} \left(v_1 \widetilde{Y}_q + v_2 Y_q
    \right) \,.
\end{equation}
To diagonalize the mass matrices we make the following unitary 
transformations on the quark fields:
\begin{subequations}
    \label{e:uni}
\begin{eqnarray}
    u'_L = V^u_L \, u_L \,, \qquad d'_L = V^d_L \, d_L \,, %\\  
  \qquad  u'_R = V^u_R \, u_R \,, \qquad d'_R = V^d_R \, d_R \,,
\end{eqnarray}
\end{subequations}
where $q'$ represents a physical quark field in the mass basis.
Now the bidiagonalization of the mass matrices can be performed
as follows:
\begin{equation}
    \label{e:bi}
    {\cal D}_u = V_L^u M_u {V_R^u}^\dagger
    = {\rm diag}\{m_u, m_c, m_t \} \,, \qquad
    {\cal D}_d = V_L^d M_d {V_R^d}^\dagger
    = {\rm diag}\{m_d, m_s, m_b \} \,.
\end{equation}
The Yukawa couplings of $h_1$ and $h_2$ (defined in \Eqn{e:SSB})
can be obtained from the Lagrangian of \Eqn{e:yuk} as follows:
\begin{eqnarray}
    \label{e:h1h2}
    \mathscr{L}^q_{h_1,h_2} = 
    -\frac{1}{\sqrt{2}}\overline{u}_L \left(
    h_1 Y_q + h_2\widetilde{Y}_q\right) u_R - \frac{1}{\sqrt{2}}\overline{d}_L
    \left(h_1 \widetilde{Y}_q + h_2 Y_q\right) d_R  \,,
\end{eqnarray}
Using the definition of \Eqn{e:h}, we can find the projections
of $h_1$ and $h_2$ onto $h$ as follows:
\begin{eqnarray}
    \label{e:ortho}
    \begin{pmatrix} h \\ h^{'} \\ h^{''} \\ h^{'''} \end{pmatrix} = 
    \frac{1}{v} \begin{pmatrix} v_1 & v_2 & v_L & 0 \\
    \cdots & \cdots & \cdots & \cdots\\ 
    \cdots & \cdots & \cdots & \cdots\\ 
\cdots & \cdots & \cdots & \cdots \end{pmatrix}
    \begin{pmatrix} h_1 \\ h_2 \\ h_L \\ h_R \end{pmatrix}
    &\Rightarrow&
    \begin{pmatrix} h_1 \\ h_2 \\ h_L \\ h_R \end{pmatrix} = 
    \frac{1}{v} \begin{pmatrix} v_1 & \cdots & \cdots & \cdots \\
    v_2 & \cdots & \cdots & \cdots\\ 
    v_L & \cdots & \cdots & \cdots\\ 
    0 & \cdots & \cdots & \cdots \end{pmatrix}
    \begin{pmatrix} h \\ h^{'} \\ h^{''} \\ h^{'''} \end{pmatrix} \,,
\end{eqnarray}
where, in the last step, we have used the fact that the transformation
matrix is orthogonal.
Now we can use this to replace $h_1$ and $h_2$ in \Eqn{e:h1h2} and extract the Yukawa couplings of $h$ as
\begin{eqnarray}
    \label{e:hyuk}
    \mathscr{L}^q_h &=& -\frac{h}{v} \overline{u}'_L
    \left[V_L^u\frac{1}{\sqrt{2}} \left(v_1 Y_q + v_2
    \widetilde{Y}_q\right){V_R^u}^\dagger\right] u'_R  
    -\frac{h}{v}
    \overline{d}'_L \left[V_L^d\frac{1}{\sqrt{2}} \left(v_2 Y_q +
    v_1\widetilde{Y}_q\right) {V_R^d}^\dagger\right] d'_R 
    +\hc \,,\nonumber\\
    &=& -\frac{h}{v} \overline{u}'_L {\cal D}_u u'_R -\frac{h}{v} 
    \overline{d}'_L {\cal D}_d \, d'_R  +\hc
  \, \, \equiv \,\, -\frac{h}{v}\left(\overline{u}' {\cal D}_u u'
   +\overline{d}' {\cal D}_d d'\right) \,.
\end{eqnarray} 
Evidently, the Yukawa couplings of $h$ are also SM-like.

%%%%%%%%%%%%%%%%%%%%%%%%%%%%%%%%%%%%%%%%%%%%%%%%%%%%%%%%%%%%%
\bibliographystyle{JHEP} 
\bibliography{DT}
\end{document}